\documentclass[conference]{IEEEtran}
\ifCLASSINFOpdf
\else
\fi
\usepackage{multirow}
\usepackage{graphicx}
\usepackage{subfigure}
\usepackage{amsmath}
\usepackage{hyperref}
\usepackage[table,xcdraw]{xcolor}
\begin{document}

\title{Cough Detection Using Selected Informative Features from Audio Signals}

\author{\IEEEauthorblockN{Xinru Chen}
\IEEEauthorblockA{School of Communication \\and Electronic Engineering\\
East China Normal University\\
E-mail:10184800230@stu.ecnu.edu.cn}
\and
\IEEEauthorblockN{Menghan Hu*}
\IEEEauthorblockA{School of Communication \\and Electronic Engineering\\
East China Normal University\\
E-mail:mhhu@ce.ecnu.edu.cn\\
*Corresponding author}
\and
\IEEEauthorblockN{Guangtao Zhai}
\IEEEauthorblockA{Institute of Image Communication\\and Network Engineering\\
Shanghai Jiao Tong University\\
E-mail:zhaiguangtao@sjtu.edu.cn}}

%\IEEEauthorblockA{\IEEEauthorrefmark{3}Starfleet Academy, San Francisco, %California 96678-2391\\
%Telephone: (800) 555--1212, Fax: (888) 555--1212}


%\IEEEauthorblockA{\IEEEauthorrefmark{1}School of Electrical and Computer %Engineering\\
%Atlanta, Georgia 30332--0250}
%\IEEEauthorblockA{\IEEEauthorrefmark{2}Twentieth Century Fox, Springfield, USA}
%\IEEEauthorblockA{\IEEEauthorrefmark{3}Starfleet Academy, San Francisco, %California 96678-2391\\
%Telephone: (800) 555--1212, Fax: (888) 555--1212}


\maketitle

\begin{abstract}
%\boldmath
Cough is a common symptom of respiratory and lung diseases. Cough detection is important to prevent, assess and control epidemic, such as COVID-19. This paper proposes a model to detect cough events from cough audio signals. The models are trained by the dataset combined ESC-50 dataset with self-recorded cough recordings. The test dataset contains inpatient cough recordings collected from inpatients of the respiratory disease department in Ruijin Hospital. We totally build 15 cough detection models based on different feature numbers selected by Random Frog, Uninformative Variable Elimination (UVE), and Variable influence on projection (VIP) algorithms respectively. The optimal model is based on 20 features selected from Mel Frequency Cepstral Coefficients (MFCC) features by UVE algorithm and classified with Support Vector Machine (SVM) linear two-class classifier. The best cough detection model realizes the accuracy, recall, precision and F1-score with 94.9\%, 97.1\%, 93.1\% and 0.95 respectively. Its excellent performance with fewer dimensionality of the feature vector shows the potential of being applied to mobile devices, such as smartphones, thus making cough detection remote and non-contact. 
\end{abstract}


% \fi
\IEEEpeerreviewmaketitle

\section{Introduction}
Cough is one of the commonest symptoms of respiratory and lung diseases such as asthma, pertussis and pneumonia. It is a body mechanism to clear upper respiratory tract and eject excessive amount of mucus and foreign particles from respiratory system \cite{1}. The mechanical phases of a cough event include that the glottis closes after deep inhalation, and the glottis opens while rapid expiratory flow occurs, producing a specific audio signal \cite{2}. 

Because cough events convey vital information of the state of the respiratory system and the status of the disease progression, various cough event assessment devices have been developed to detect cough events and calculate cough frequency, cough intensity, cough impact, cough duration and other indicators \cite{3}. In addition to manual assessment methods such as manual cough counter, objective cough assessment devices such as the Leicester Cough Monitor \cite{4}, the Hull Automatic Cough Counter \cite{5} and VitaloJak system \cite{6} are available. Although these objective cough assessment devices can be used as wearable cough detection systems with high accuracy, they are not suitable for large-scale cough screening applications. 
 
Cough event detection for a large number of people plays an important role in epidemic prevention and control, epidemiological research, risk assessment of infectious diseases and other fields. For example, by July 30 2021, COVID-19, a global pandemic, has confirmed 196,553,009 cases and caused 4,200,412 deaths, according to World Health Organization (WHO) \cite{7}. Due to the spread of the mutant COVID-19 coronavirus and the unbalanced vaccine supply, the global COVID-19 pandemic is facing a severe situation again. In the COVID-19 pandemic, the primary symptom of a patient with COVID-19 coronavirus may be cough \cite{8}. Therefore, large-scale and contactless cough detecting applications are of great importance. 

The occurrence of cough events is usually accompanied by the production of some specific sounds. Therefore, many scholars used audio signals to build models, so as to achieve remote detection of cough events. Alsabek reported the importance of Mel Frequency Cepstral Coefficients (MFCC) extraction of COVID-19 and non-COVID-19 samples in cough signal process \cite{9}. Monge-Álvarez reported that using local Hu moments as a robust feature for cough detection with audio signals \cite{10}. Al-Khassaweneh used Wigner distribution and wavelet transform to analyze cough signals for detecting asthma \cite{11}. Pramono developed an algorithm based on features of the audio signals with a logistic regression model to identify cough events \cite{12}. Laguarta developed an 
Artificial Intelligence (AI) speech processing architecture based on a Convolutional Neural Network (CNN) to screening the coughing symptom of COVID-19 \cite{13}. Imran proposed a COVID-19 diagnosis system using the combination of machine learning and deep learning from cough samples via an App \cite{24}.

Most studies we mentioned above extract a large number of feature and classify with complex deep learning or machine learning model, which may be infeasible or time-consuming to apply in remote mobile devices, such as smartphones. Meanwhile, it may increase the risk of overfitting. Cough detection, a relatively simple problem, does not require complicated technical solutions to solve. What's more, most studies do not include cough samples recorded in scenarios where cough detection is required such as the ward in their dataset. Therefore, it is necessary to develop a cough detection model using small number of features and machine learning algorithm such as Support Vector Machine (SVM) while achieving optimal performance with inpatient cough samples. 

SVM, a well-known classifier, is widely used in various detection and identification techniques. Kumar presented a person authentication framework using SVM classifier to implement person identification and verification \cite{20}. Song proposed a sound-of-tapping technology based on SVM, including 7 tapping position classification and 6 medium identification \cite{21}. Many studies used SVM as a classifier to implement cough detection. Bhateja reported a method for features extraction and classification of cough sound in noisy environment using SVM \cite{22}. Vhaduri developed a cough and snore detection framework based on several machine learning classifiers including SVM, which can be implemented on a smartphone \cite{23}.

In this paper, we compare the performance of models using different feature space optimization methods, including Random Frog, Uninformative Variable Elimination (UVE) and Variable influence on projection (VIP), and propose the optimal model of automatic cough detection which uses 20 features selected from MFCC features by UVE algorithm and classifies with SVM linear two-class classifier. Our model can achieve 94.9\% accuracy of classifying between cough audios and non-cough audios. 

In Section 2, we summarize the methods and the specific process of our cough detection model. Section 3 explains the dataset we used to train and test our model, describes the experimental results with performance metrics and discusses the potential and limitations of our study. Section 4 draws a conclusion.

\section{Method}
\subsection{Feature Extraction and Dimension Reduction}
We first extract Mel Frequency Cepstral Coefficients (MFCC) from the cough audios result in a $5000\times36$ feature matrix. Then, we take the top few Principal Component Analysis (PCA) projections of the MFCC features which keep 95\% of the main information and combine them into a feature vector with the dimension of 107. By combining MFCC feature extraction method and PCA, it was expected to improve the accuracy in automatic speech recognition system and reduce the feature dimension \cite{14}.
\subsubsection{MFCC}

For automatic speech recognition, MFCC features have been found to a useful feature extraction method \cite{14}. The Mel scale describes the nonlinear characteristics of human ear frequency, and its relationship with frequency can be approximated by the following equation:
\begin{equation}
\operatorname{Mel}(f)=2595 \times \lg \left(1+\frac{f}{700}\right)
\end{equation}
It is aimed to reflect audible changes and conform to the auditory characteristics of the human ear more closely when making changes in frequency. Then, the Cepstral analysis is performed on the Mel spectrum of cough audios to compute their Cepstral coefficients, thus obtaining MFCC features. The process of MFCC is shown in Fig. 1(a). 
\begin{figure}[ht]
\centering
\subfigure[MFCC diagram block]{
\begin{minipage}[t]{0.5\linewidth}
\centering
\includegraphics[scale=0.6]{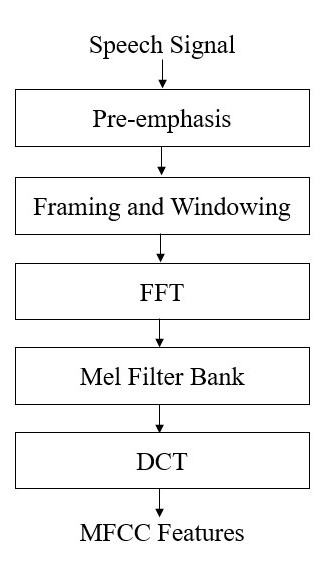}
%\caption{fig1}
\end{minipage}%
}%
\subfigure[PCA diagram block]{
\begin{minipage}[t]{0.5\linewidth}
\centering
\includegraphics[scale=0.6]{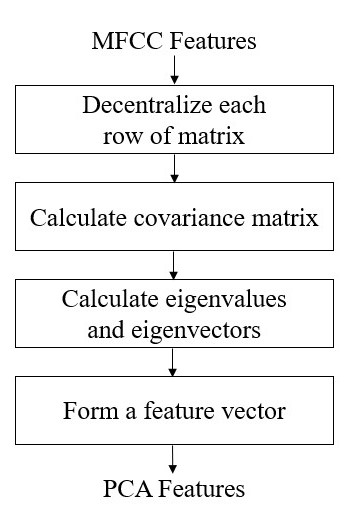}
%\caption{fig2}
\end{minipage}%
}%

\caption{Diagram block of feature extraction (a) and dimension reduction (b)}
\label{fig:label}
\end{figure}

\subsubsection{PCA}

PCA is a useful dimension reduction method with high correlation level. PCA processing steps can be seen in Fig. 1(b).

\paragraph{}
There are $m$ samples with $n$ dimension to form a $n \times m$ matrix $X$. Decentralize each row of $X$ by subtracting the mean of that row.

\paragraph{}
Calculate the covariance matrix $C=\frac{1}{m} X X^{\top}$.  

\paragraph{}
Calculate the eigenvalues and corresponding eigenvectors of the covariance matrix.

\paragraph{}
Form a feature vector. The eigenvectors are arranged in rows from top to bottom according to the size of corresponding eigenvalues, and the first $k$ rows are taken to form matrix $P$.

\paragraph{}
$Y=P X$ is the feature vector after dimensionality reduction to $k$.

\subsection{Feature Space Optimization}

To select informative features, reduce the complexity of the model and the risk of overfitting, and make future practical application more effective at the same time, we use the following three feature selection algorithms: Random Frog, UVE algorithm and VIP algorithm.

\subsubsection{Random Frog}

Random Frog is an effective feature selection algorithm for high dimensional feature. It can use a small number of feature iteration for modeling and output the possibility of each feature selection, which can be used as a feature selection criterion \cite{15}.
Fig. 2 shows the key steps of Random Frog algorithm. Its main calculation steps include the following three steps:

\paragraph{}
Given an initial variable subset $V_{0}$ and initialize $V_{0}$ with $Q$ variables. 

\paragraph{}
Based on the initial variable subset $V_{0}$, a candidate variable subset $V^{*}$, including $Q^{*}$ variables, is proposed. Choose $V^{*}$ as $V_{1}$ instead of the initial variable subset $V_{0}$. Repeat the above process until $N$ iterations are finished. 

\paragraph{}
Compute the selection probability of each variable and use the selection probability as the criterion for selecting variables. 

\begin{figure*}[ht]
\centering
\includegraphics[scale=0.6]{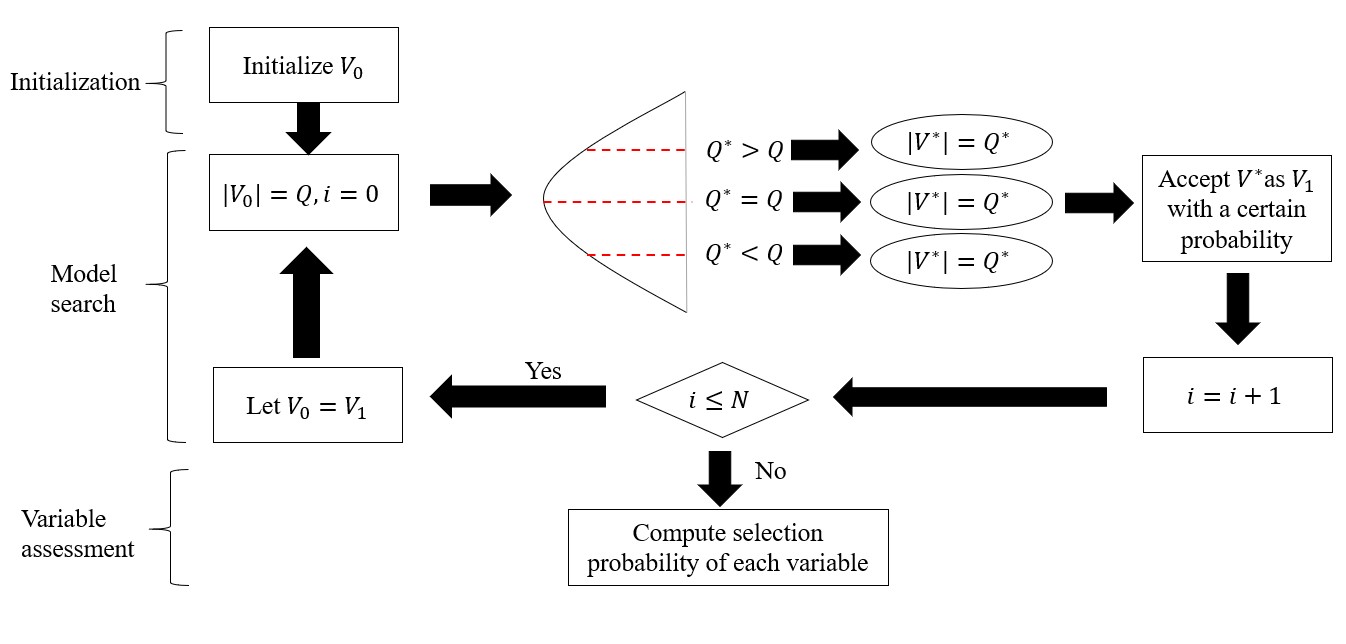}
\caption{Key steps of Random Frog algorithm. $V_{0}$: Variable subset; $Q$: Number of variable in $V_{0}$; $V^{*}$: Candidate variable subset; $Q^{*}$: Number of variable in $V^{*}$; $i$: Iterations. (Note: From reference \cite{15}.)}
\label{fig:label}
\end{figure*}

Based on the single $107 \times 1$ feature vector obtained from PCA, we select 10, 20, 30, 40 and 50 features respectively using Random Frog algorithm.

\subsubsection{UVE Algorithm}
UVE is a feature selection algorithm based on a stability analysis of PLS regression coefficients which can eliminate variables of invalid information in the PLS regression model, thus improving the speed, anti-interference ability, and predictive stability of PLS models \cite{16}. UVE algorithm’s main principle is to add artificial random variables to the original variable matrix, calculate a selection criterion for original and random variables, and keep original variables whose selection criterion result is larger than that of random variables. The selection criterion calculation is the ratio of regression coefficient $b_{j}$ and its standard deviation $\boldsymbol{s}\left(b_{j}\right)$, measuring the reliability of the PLS regression coefficient $c_{j}$ \cite{17}.
\begin{equation}
c_{j}=b_{j} / s\left(b_{j}\right) \text { for } j=1, \ldots, p
\end{equation}

We select 10, 20, 30 and 40 features respectively from the single $107\times1$ feature vector obtained from PCA using UVE algorithm.

\subsubsection{VIP Algorithm}
VIP algorithm is used to screen out the most relevant variables. VIP is a parameter that calculates the cumulative influence of individual X-variables in a PLS model. Equation 3 gives a detailed calculation of VIP:
\begin{equation}
V I P_{P L S}=\sqrt{K \times\left(\frac{\left[\Sigma_{a=1}^{A}\left(W_{a}^{2} \times S S Y_{\text {comp }, a}\right)\right]}{S S Y_{\text {cum }}}\right)}
\end{equation}
where a means PLS dimension, $K$ means the total number of variables, $W_{a}^{2}$ means the squared PLS weight, and $SS$ means the explained the sum of squares \cite{18}. The values of VIP above 1 are considered as the most relevant variables, and the values of VIP smaller than 0.5 express irrelevant variables. We select top 10, 20, 30, 40 and 50 features whose VIP values are larger than 1 according to the single $107\times1$ feature vector obtained from PCA.

\subsection{Cough and Non-cough Classification}
On the basis of the features selected, we use classifiers to classify cough samples and non-cough samples, thus detecting cough. In the current work, SVM linear dichotomous classifiers is used for the classification. The basic idea of SVM is to find the separation hyperplane that can divide the training data set correctly and maximize the margin between the two classes. SVM constructs a hyperplane in a high dimensional space which can be used for classification, and training data samples are referred as points in the high dimensional space. The points that are nearest to the hyperplane drawn previously are known as Support Vectors, and the distance between these vectors and the hyperplane is defined as margin. SVM model attempts to find the optimal hyperplane with maximum margin distance. 

\section{Experiments}
\subsection{Dataset Explanation}

Our dataset consists of cough samples and non-cough samples. The cough samples include cough sounds in Environmental Sound Classification (ESC-50) dataset, self-recorded cough recordings and inpatient cough recordings from the patients with respiratory diseases. The non-cough samples are chosen from labeled environmental recordings in ESC-50 dataset, including animal sounds, natural sounds, human (non-speech) sounds, interior sounds and other exterior noises \cite{19}. The inpatient cough recordings are collected from inpatients of the respiratory disease department in Ruijin Hospital. The inpatients we recorded suffered from respiratory disease and had symptoms of cough. The cough samples collected were recorded by mobile microphone, and background noise is included in these cough samples. The duration of every sample in our dataset is 5 seconds. 
\begin{figure*}[h]
\centering
\subfigure[40 features selected by Random Frog]{
\begin{minipage}[t]{\textwidth}
\centering
\includegraphics[width=6in]{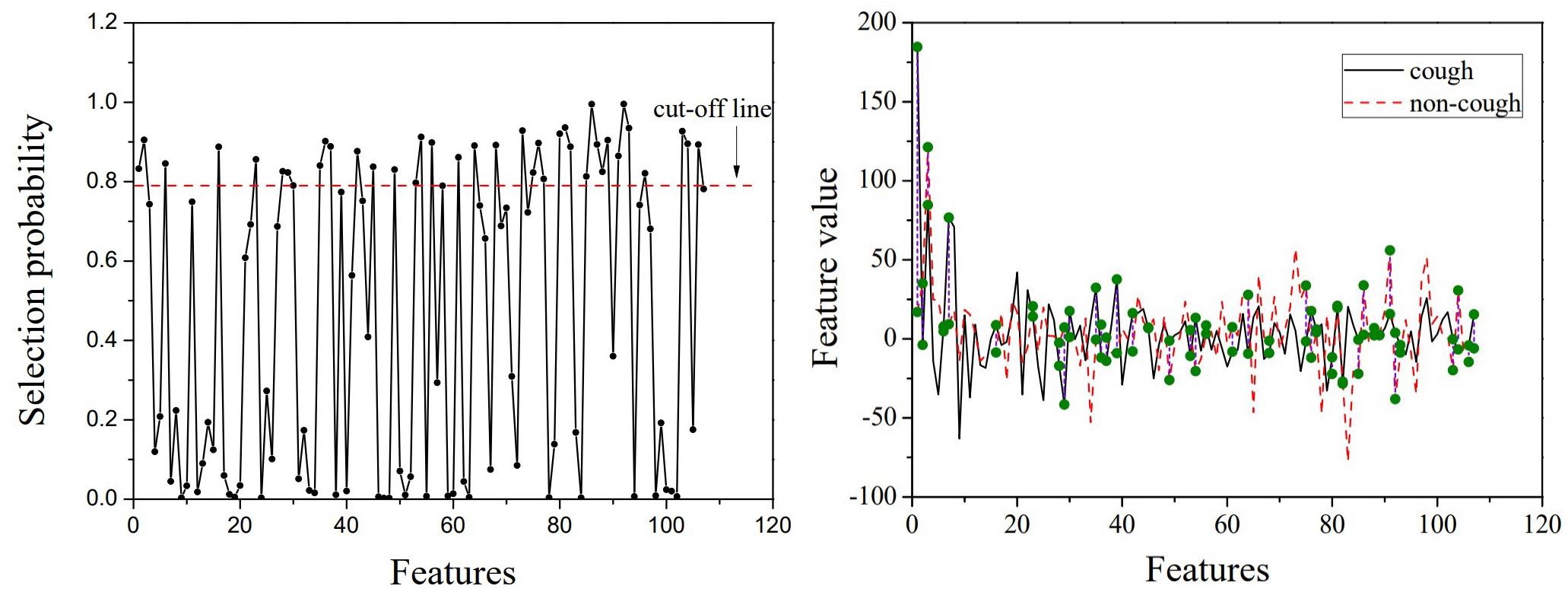}
%\caption{fig1}
\end{minipage}%
}%

\subfigure[20 features selected by UVE algorithm]{
\begin{minipage}[t]{\textwidth}
\centering
\includegraphics[width=6in]{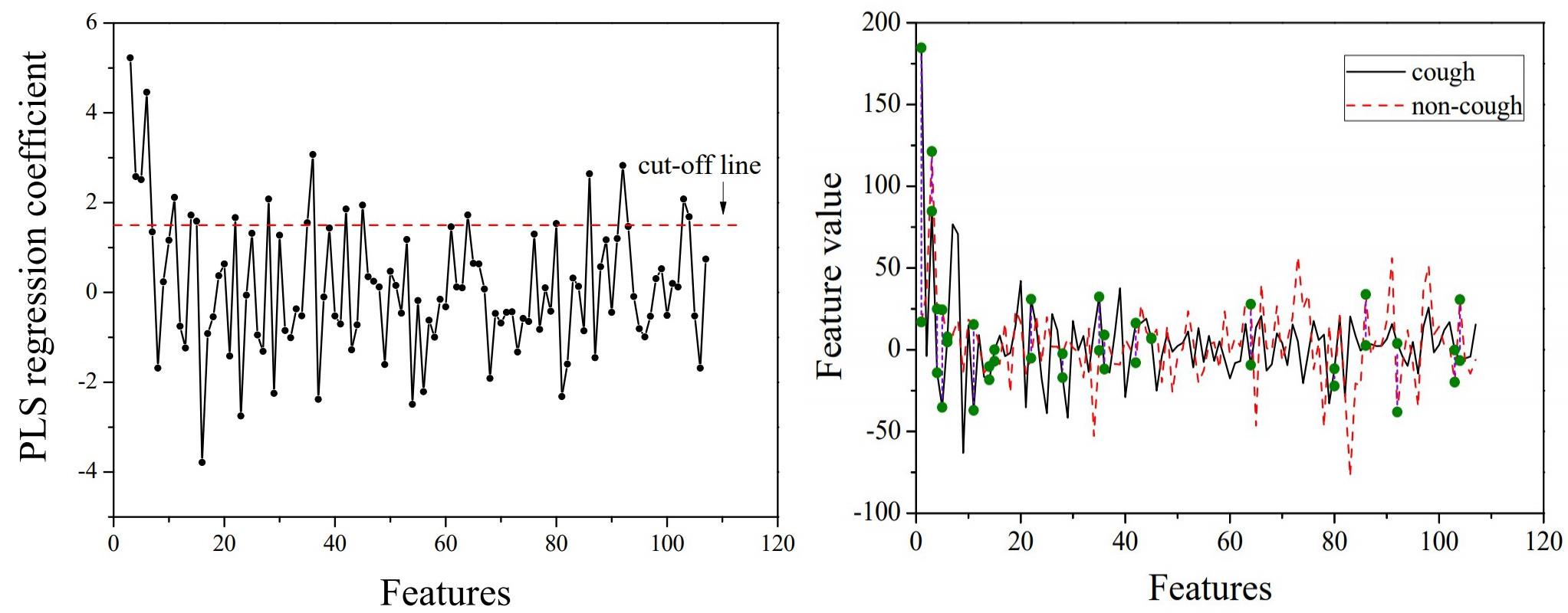}
%\caption{fig2}
\end{minipage}%
}%

\subfigure[10 features selected by VIP algorithm]{
\begin{minipage}[t]{\textwidth}
\centering
\includegraphics[width=6in]{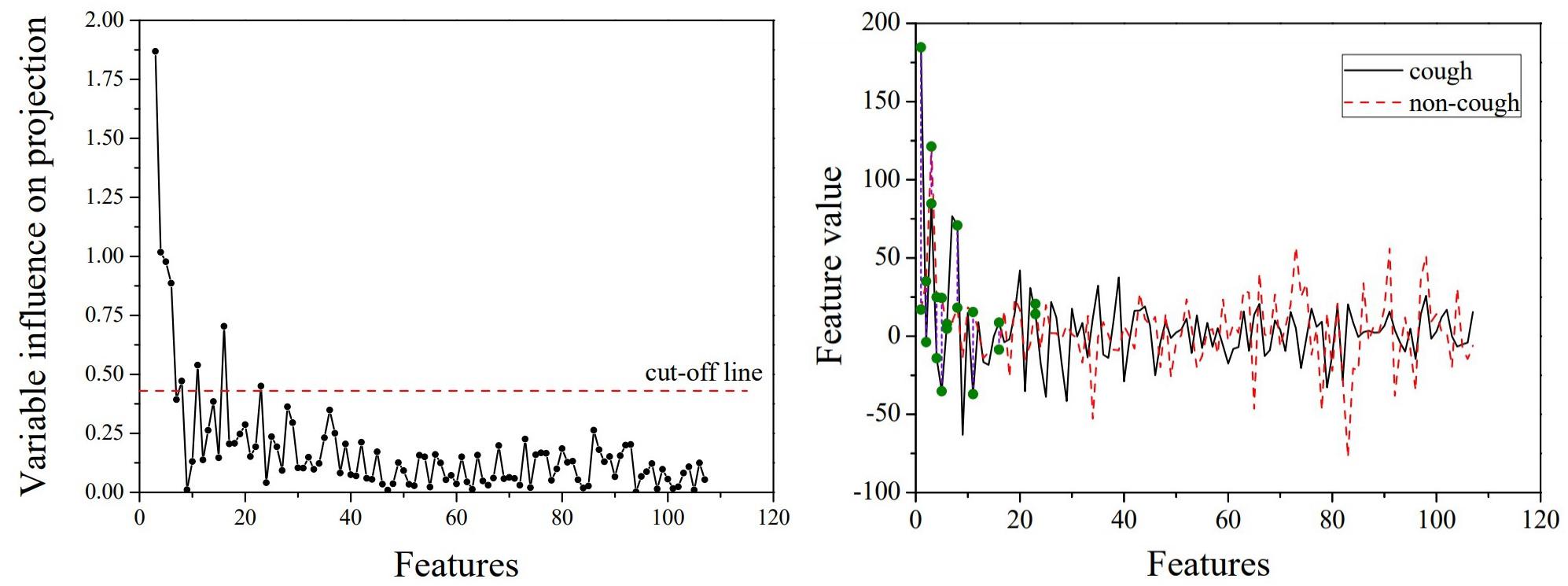}
%\caption{fig2}
\end{minipage}%
}%
\caption{Importance of features from three feature selection algorithms (left column) and Differences between feature value of a typical cough audio and a non-cough audio (right column)}
\label{fig:label}
\end{figure*}

Finally, we obtained 335 cough samples and 335 non-cough samples, a total of 670 samples. In the classification process, we separate our dataset into training dataset and testing dataset. The training dataset contains 266 cough samples and 266 non-cough samples.The test dataset includes 69 cough samples and 69 non-cough samples. All the cough samples in the test dataset are inpatient cough recordings collected from inpatients with respiratory diseases. 

\subsection{Informative Features Selection}
We choose the best model for each feature selection algorithm introduced in Section 2 and draw the importance of features in Fig. 3 (left column). The importance of features from Random Frog, UVE and VIP is respectively described as selection probability, PLS regression coefficient and variable influence on projection. The features with importance above the cut-off threshold drawn in Fig. 3 (left column) are chosen as informative features. Fig. 3 (right column) depicts the feature value of a typical cough audio and a typical non-cough audio after PCA processing. The informative features selected by three feature selection algorithms are respectively marked with green dots in the right column of Fig. 3 (a-c). In addition, Fig. 3 (right column) shows the differences between feature value of the cough audio and the non-cough audio with red dotted lines, indicating that the selected informative features are more effective for classification than the original features. Six features overlap in three models shown in Fig. 3 (right column) despite using different feature selection algorithms, which means three algorithms can select audio features informative to cough.

\begin{table*}[h]
\begin{center}
\textbf{Table 2}~~: Performance metrics for different feature selection methods. (The bold rows are the top five performing models. The row with gray background 
\leftline{is the best performing model.)}
\resizebox{\textwidth}{!}{
\begin{tabular}{c|l|l|l|l|l}
\hline
\multicolumn{1}{l|}{Feature selection method} & Feature number                      & Accuracy (\%)                          & Sensitivity/Recall (\%)               & Precision (\%)                         & F1-Score                              \\ \hline
PCA                                           & 107                                 & 85.51                                  & 100                                   & 77.53                                  & 0.87                                  \\ \hline
                                              & 10                                  & 67.39                                  & 100                                   & 60.53                                  & 0.75                                  \\
                                              & 20                                  & 79.71                                  & 100                                   & 71.13                                  & 0.83                                  \\
                                              & 30                                  & 85.51                                  & 100                                   & 77.53                                  & 0.87                                  \\
                                              & 40                                  & 86.23                                  & 100                                   & 78.41                                  & 0.88                                  \\
\multirow{-5}{*}{Random Frog}                 & 50                                  & 81.16                                  & 100                                   & 72.63                                  & 0.84                                  \\ \hline
                                              & 10                                  & 28.99                                  & 7.25                                  & 12.82                                  & 0.09                                  \\
                                              & \cellcolor[HTML]{C0C0C0}\textbf{20} & \cellcolor[HTML]{C0C0C0}\textbf{94.93} & \cellcolor[HTML]{C0C0C0}\textbf{97.1} & \cellcolor[HTML]{C0C0C0}\textbf{93.06} & \cellcolor[HTML]{C0C0C0}\textbf{0.95} \\
                                              & \textbf{30}                         & \textbf{94.20}                         & \textbf{95.65}                        & \textbf{92.96}                         & \textbf{0.94}                         \\
\multirow{-4}{*}{UVE}                         & \textbf{40}                         & \textbf{92.03}                         & \textbf{97.1}                         & \textbf{88.16}                         & \textbf{0.92}                         \\ \hline
                                              & \textbf{10}                         & \textbf{90.58}                         & \textbf{100}                          & \textbf{84.15}                         & \textbf{0.91}                         \\
                                              & \textbf{20}                         & \textbf{87.68}                         & \textbf{100}                          & \textbf{80.23}                         & \textbf{0.89}                         \\
                                              & 30                                  & 84.06                                  & 100                                   & 75.82                                  & 0.86                                  \\
                                              & 40                                  & 85.51                                  & 100                                   & 77.53                                  & 0.87                                  \\
\multirow{-5}{*}{VIP}                         & 50                                  & 75.36                                  & 100                                   & 66.99                                  & 0.80                                  \\ \hline
\end{tabular}}
\end{center}
\end{table*}

\subsection{Classification Result}
\begin{table}[]
\begin{center}
\textbf{Table 1}~~: Confusion matrix for cough detection.\\

\begin{tabular}{cc|c|c|}
\cline{3-4}
                                                  &           & \multicolumn{2}{c|}{Predicted Class}    \\ \cline{3-4} 
                                                  &           & Cough              & Non-cough          \\ \hline
\multicolumn{1}{|c|}{\multirow{2}{*}{True Class}} & Cough     & True Positive(TP)  & False Negative(FN) \\ \cline{2-4} 
\multicolumn{1}{|c|}{}                            & Non-cough & False Positive(FP) & True Negative(TN)  \\ \hline
\end{tabular}
\end{center}
\end{table}

To evaluate the performance of models, we use the confusion matrix in Table 1 and calculate the performance metrics of accuracy, precision, sensitivity/recall, and F1-score on the test dataset. 
\begin{equation}
\text { accuracy }=\frac{T P+T N}{T P+T N+F P+F N}
\end{equation}
\begin{equation}
\text { precision }=\frac{T P}{T P+F P}
\end{equation}
\begin{equation}
\text { sensitivity/recall }=\frac{T P}{T P+F N}
\end{equation}
\begin{equation}
\text { F1-score }=2 *\left(\frac{\text { precision } \times \text { recall }}{\text { precision }+\text { recall }}\right)
\end{equation}

We first use the single $107\times1$ feature vector obtained from the dimension reduction of PCA to train SVM classifier. Then, we respectively use 10, 20, 30, 40 and 50 features selected by Random Frog, UVE and VIP. We totally build 15 cough detection models and evaluate their performance of detecting cough events via the test dataset. The classification results of these models are presented in Table 2.

According to the results in Table 2, we can see that the model with 20 features selected by UVE algorithm can achieve the best performance, allowing classification with accuracy, recall, precision and F1-score of 94.9\%, 97.1\%, 93.1\%, and 0.95, respectively. The results indicate that our cough detection model can reach 94.9\% accuracy of classifying between cough event and non-cough event and use fewer dimensionality of the feature vector. 

\subsection{Discussion}
The model with 20 features selected by UVE algorithm shows that the classification performance can be optimized while reducing dimension of the feature vector. Our model has low complexity, so it is well suited for deployment at the edge such as smartphones. In addition, the use of audio combined with our low-complexity model minimizes the computational resources at the edge. Therefore, it is ideal for long-term deployment in public places such as wards and subways to monitor the frequency of coughs and thus providing decision support for inspection and quarantine. 

The performance of our cough detection model is limited by the following factors:

\paragraph{Inadequate ambient noise of the training samples}
The training samples we used were recorded in a quiet environment using a stationary mobile microphone, while the test samples have ambient noise. We will try recording cough and non-cough samples in a more noisy and less controlled environment and changing the microphone position during recording, which is more consistent with the actual cough detection applied to mobile devices. 

\paragraph{Quantity of the training and test samples}
Our cough dataset is not so big enough. In the ongoing work, we will enlarge our cough recording dataset and combine deep learning algorithm to optimize our cough detection model. 

\paragraph{Extraction method of features}
Our model only uses selected MFCC features. In the future work, we can combine MFCC with LPC coefficient and spectral features, such as spectral flatness and spectral centroid.

\section{Conclusion}
We have developed a cough detection model using SVM classifier and selected informative features. To train our model, we have used a cough and non-cough dataset combined ESC-50 dataset with cough samples recorded from people around our authors using mobile microphone. The test dataset we used is inpatient cough recordings collected from inpatients of the respiratory disease department in Ruijin Hospital, who had symptoms of cough. We first extract MFCC features from the audios in dataset, take the top PCA projections of the MFCC features to reduce dimension, and combine them into a single $107\times1$ feature vector. Then, we use three feature selection algorithms, namely Random Frog, UVE, and VIP, and build several models with different numbers of features. Finally, we use SVM linear two-class classifier to classify cough samples and non-cough samples. We have found that the model with 20 features selected by UVE algorithm can achieve the best model performance with 94.9\% accuracy, 97.1\% recall, 93.1\% precision and the F1-score of 0.95, respectively. 

%Although our model needs to be further validated on a larger dataset, the model performance we have presented shows that our cough detection model is viable based on the classification using a small number of cough audio features. Furthermore, it’s feasible for our model to be applied, collect cough data and classify in mobile devices remotely, such as smartphones, thus making cough detection more cost-efficient and contactless. This cough detection tool may decrease the burden on global health system and reduce contact between health care workers and patients. 

In the future work, we will enlarge our dataset, make our model more noise-proof and improve the performance of our models. We will also try to implement our cough detection model on customers’ smartphones via an App.


\section*{Acknowledgment}
This work is sponsored by the National Natural Science Foundation of China (No. 61831015, No. 61901172) , the Shanghai Sailing Program (No. 19YF1414100), and the Science and Technology Commission of Shanghai Municipality (No. 19511120100, No. 18DZ2270700, No. 18DZ2270800).  

%We would like to thank the 



%\bibliographystyle{IEEEtran}
%\begin{thebibliography}{1}

%H.~Kopka and P.~W. Daly, \emph{A Guide to \LaTeX}, 3rd~ed.\hskip 1em %plus

%\end{thebibliography}
%\printbibliography{}
\newcommand\BIBentryALTinterwordstretchfactor{2.5}
\bibliographystyle{IEEEtran}
\bibliography{IEEEabrv,cough}

\end{document}